\documentclass{nature}

\usepackage{subfigure}
\usepackage{array}
\usepackage{float}

\usepackage{graphicx}
\usepackage{amsmath}
\usepackage{amssymb}
\usepackage{lineno}
\usepackage{caption}
\usepackage{booktabs}
\usepackage{multirow}
\usepackage{cite}
\usepackage{times}
\usepackage{color}
\usepackage{caption}
\usepackage{xurl}

\usepackage{docmute}
\usepackage[dvipsnames]{xcolor}

\title{\textbf{Universal expansion of human mobility across urban scales}}

\author{Lu Zhong${}^{1,2,6}$, Lei Dong${}^{3,6}$, Qi R. Wang${}^{4}$, Chaoming Song${}^{5}$, Jianxi Gao${}^{1,2*}$}
\begin{document}

\maketitle

\begin{affiliations}
\item{Department of Computer Science, Rensselaer Polytechnic Institute, Troy, NY, USA}
\item{Network Science and Technology Center, Rensselaer Polytechnic Institute, Troy, NY, USA}
\item{Institute of Remote Sensing and Geographical Information Systems, School of Earth and Space Sciences, Peking University, Beijing, China}
\item{Department of Civil and Environmental Engineering, Northeastern University, Boston, MA, USA} 
\item{Department of Physics, University of Miami, Coral Gables, FL, USA}
\item{These authors contributed equally: Lu Zhong, Lei Dong.}\\
Email:gaoj8@rpi.edu
\end{affiliations}

\begin{abstract} 

Human mobility is a fundamental process underpinning socioeconomic life and urban structure. Classic theories, such as egocentric activity spaces and central place theory, provide crucial insights into specific facets of movement, like home-centricity and hierarchical spatial organization. However, identifying universal characteristics or an underlying principle that quantitatively links these disparate perspectives has remained a challenge. Here, we reveal such a connection by analyzing the spatial structure of individual daily mobility trajectories using network-based modules. We discover a universal scaling law: the spatial extent (radius) of these mobility modules expands sublinearly with increasing distance from home, a pattern consistent across three orders of magnitude. Furthermore, we demonstrate that these modules precisely map onto the nested hierarchy of urban systems, corresponding to local, city-level, and regional scales as distance from home increases. These findings deepen our understanding of human mobility dynamics and demonstrate the profound connection between classical urban theory, human geography, and mobility studies.

\end{abstract}
In human societies, the dynamic and diverse movement of individuals underpins nearly all aspects of socioeconomic life---from the formation of social ties to the spread of disease---and is essential to the development of infrastructure and urban amenities\cite{gonzalez2008understanding,belik2011natural}. Over the past century, several seminal theories have profoundly influenced our understanding of human movement\cite{barbosa2018human} (Table S1). For instance, the gravity model, rooted in spatial interaction theory, elegantly describes how movement decays with distance, capturing the inherent friction imposed by geographic space\cite{fotheringham1989spatial,roy2004spatial}. Complementing this perspective, the concept of egocentric activity spaces emphasizes the home as a crucial anchor, around which individuals organize their daily routines and movements\cite{bhat1999activity,miller2021activity}. Broadening this spatial lens, central place theory elucidates the hierarchical organization of urban systems, revealing how human activities, both near and far from home, are supported by the spatial arrangement of cities and regions\cite{batty2006hierarchy,mulligan2012central,osth2021hierarchy,van2018christaller,alessandretti2020scales}. These classic theories, though developed independently, collectively underscore the fundamental influence of an individual's home location on their mobility patterns. However, despite their individual contributions, a unifying framework that reveals the interconnectedness of these theories and captures the universal characteristics of human movement across multiple spatial scales remains elusive. To address this gap, we introduce a network-based approach designed to capture the spatial preferences inherent in both distance and trajectory of individual mobility\cite{barthelemy2011spatial}. By decomposing these networks into modules and analyzing their properties in relation to the distance from home, we aim to uncover a latent, universal pattern that underlies these seemingly disparate classic theories of movement.

\begin{figure*}[t!]
\centering
\includegraphics[scale=0.7]{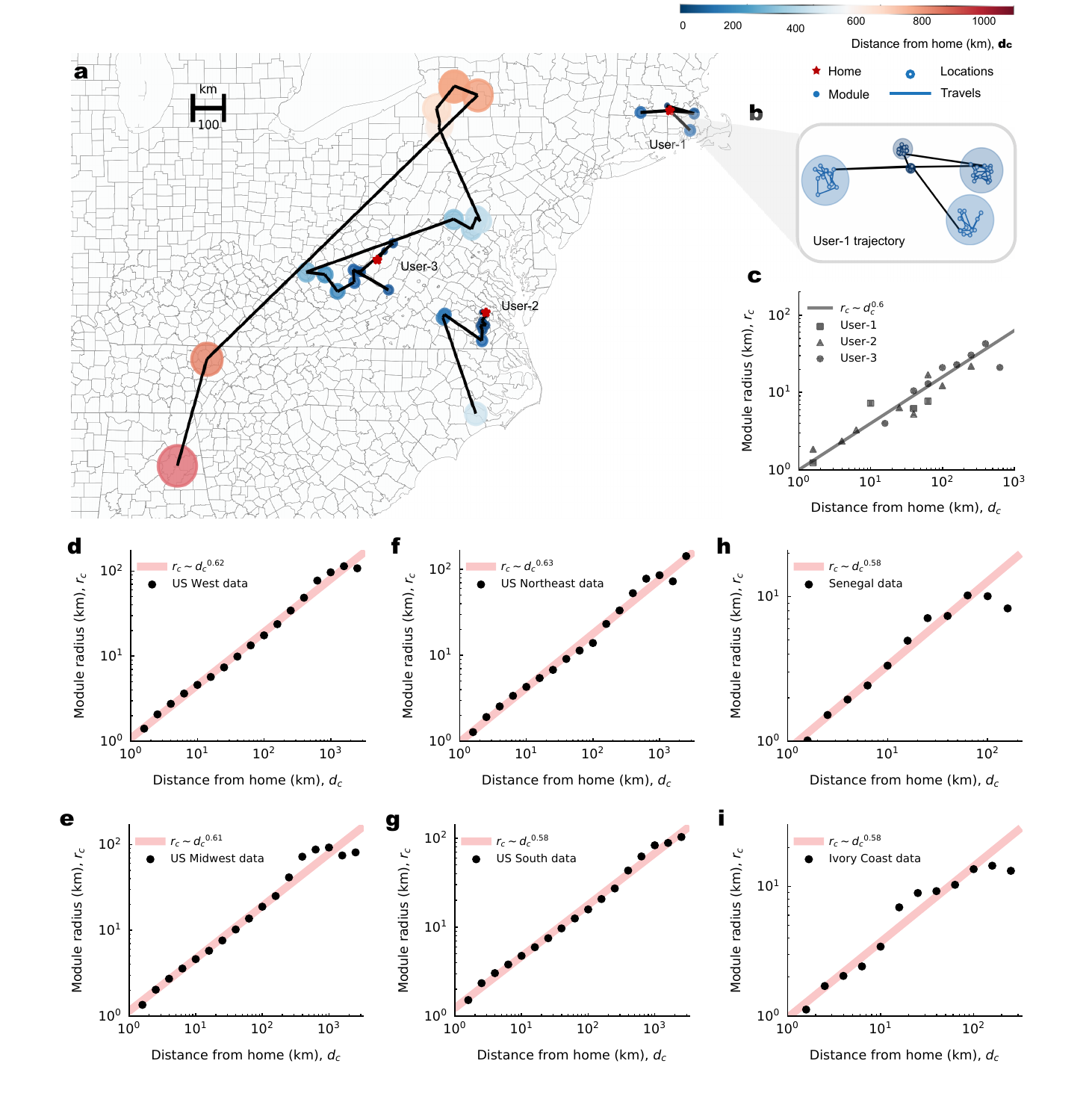}
\caption{\fontsize{10}{13.8}\linespread{1}\selectfont{\textbf{The spatial expansion of modules.} (\textbf{a}) The spatial distribution of modules in anonymized cell phone users' trajectories. By representing trajectories as networks and applying
the Louvain method, (\textbf{b}) the example trajectory network is divided into five modules, each encompassing spatially and topologically proximate locations.  (\textbf{c}) Modules located far from home are in larger spatial coverage. By analyzing all U.S., Senegal, and Ivory Coast data, \textbf{(d-i)} module radius $r_c$ increases with distance from home $d_c$ in a power-law manner, $r_c \sim d_c^{\kappa}$. Specifically, for U.S. data in the West, Northeast, Midwest, and South regions, the value of $\kappa$ is approximately $0.61 \pm 0.03$. For Senegal and Ivory Coast data, $\kappa$ is approximately $0.58$.}}
\label{figure_1}
\end{figure*}

Our analysis is based on high-resolution cell phone datasets from three countries with diverse cultural and developmental contexts (Fig. S1 and Tables S2-S3). The first dataset encompasses six months of privacy-enhanced global positioning system (GPS) trajectories from two million anonymous users in the United States (U.S.) The second and third datasets consist of two weeks of call detail records (CDR) from 300,000 anonymous users in Senegal and 50,000 anonymous users in Ivory Coast, respectively. We construct a network for each user’s daily trajectory (Fig. \ref{figure_1}a-b). In this network, stay points are modeled as nodes and the weight of edges is determined by the reciprocal of the spatial distance between stay points, reflecting inverse distance weighting principles in spatial analysis\cite{miller2004tobler}. This means that shorter spatial distances yield larger edge weights (see Methods). We apply the Louvain community detection method\cite{blondel2008fast} to partition trajectory networks into multiple modules and examine the correlation between module characteristics and their distance from the user's home (Methods, Figs. S2-S5). We focus on distance from home because numerous studies have shown that home location serves as a critical anchor in shaping human mobility patterns\cite{bhat1999activity,miller2021activity}. Figure \ref{figure_1}b illustrates a user's trajectory partitioned into five modules based on spatial and topological proximity. For each module, we measure its spatial size $r_c$ (the average distance of stay points to the module centroid) and the distance from its centroid to the home location $d_c$. 

As demonstrated by the three sample users in Fig. \ref{figure_1}c, the radius of the module increases with distance from home. Our thorough analysis of all trajectories reveals a remarkably universal pattern:
\begin{equation}
r_c \sim d_c^{\kappa}, 
\label{eq_rc}
\end{equation}
where $\kappa$ is approximately 0.61 $\pm 0.3$ for US data across the West, Northeast, Midwest, and South regions (Fig. \ref{figure_1}d-g), with $\kappa$ around 0.58 for Senegal (Fig. \ref{figure_1}h) and 0.58 for Ivory Coast (Fig. \ref{figure_1}i). We refer to this pattern in human mobility as the \textit{``expansion law"} -- as individuals move farther from home, their exploration scope (module sizes) increases sub-linearly. Specifically, when travel distances increase from $10$km, $100$km, and $1000$km, the module radius expands to $4$km, $15$km, and $63$km, corresponding to neighborhood, city, and state scales. 

To verify the generality of this expansion pattern, we conducted robustness checks on module size measurement, module extraction methods,  data sampling ratio, and demographic factors (Figs. S6-S11). We quantified module size using the convex hull of visited locations within each module (Extended Data Fig. 1) and applied alternative partitioning methods for module extraction in trajectory networks (Fig. S6). Furthermore, we estimated Eq. (1) for each U.S. state (Extended Data Figs. 2-3) and categorized users based on demographic attributes such as age, gender, race, poverty level, household income, and home location (Extended Data Fig. 4). The results demonstrate that the expansion law is robust across datasets from different regions and remains consistent under varying measures of spatial module extent, extraction methods, and diverse demographic groups. 

\begin{figure*}[t!]
\centering
\includegraphics[scale=0.85]{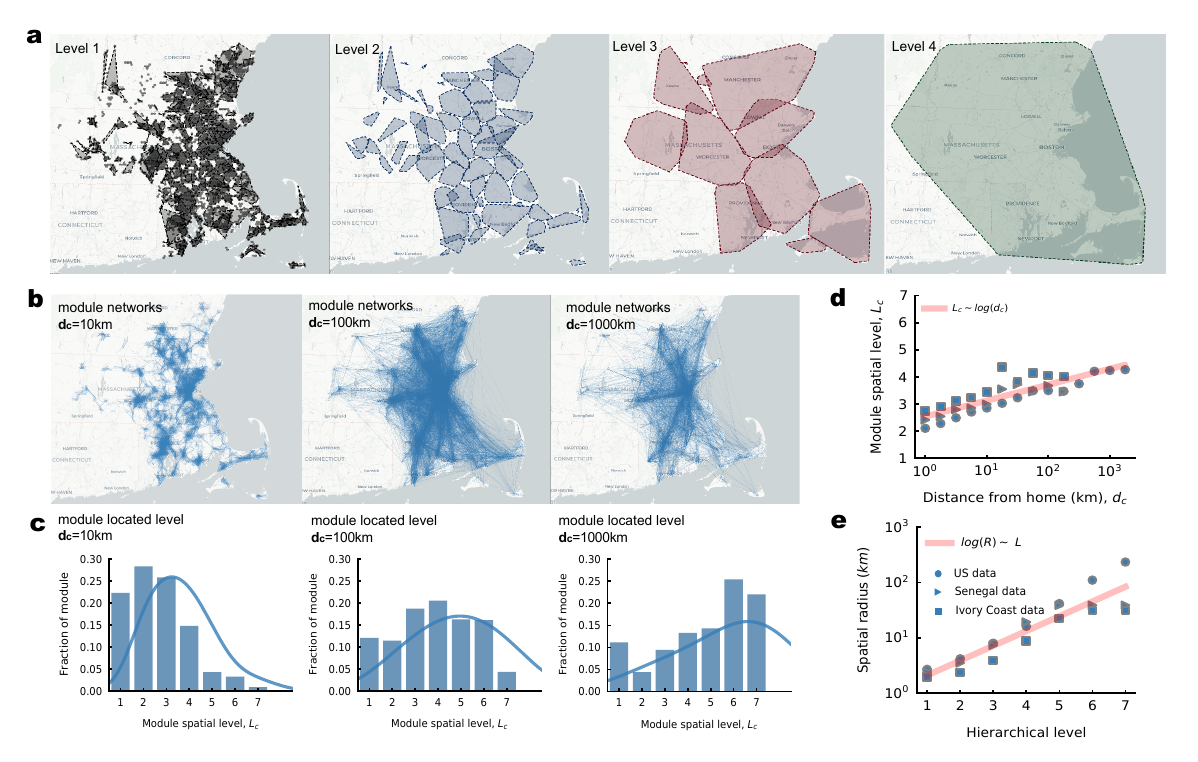}
\caption{\fontsize{10}{13.8}\linespread{1}\selectfont{\textbf{Spatial hierarchical levels and expansion law.} With the destination Boston, Massachusetts as the
example, (\textbf{a}) the urban environment is structured into four hierarchical levels $L$, with higher-level units covering lower-level units. We add the convex hull of aggregated H3 cells at each level to better illustrate the geographical extent. (\textbf{b}) Users’ module networks at the destination with varying distances from home (e.g., 10km, 100km, 1000km).  We assign modules a specific level $L_c$ if that hierarchical level unit encompasses at least 80\% of the module. \textbf{(c)} The distribution of spatial levels $L_c$ for module networks in (\textbf{b}). Modules that are at greater distances from home tend to travel across higher hierarchical levels. By analyzing all U.S., Senegal, and Ivory Coast data, \textbf{(d)} the spatial levels of modules at varying distances from home follow $L_c \sim \log(d_c)$. (\textbf{e}) The spatial size of hierarchical levels follows $\log(R) \sim L$, leading to $\log(R) \sim L=L_c \sim \log(d_c)$. }}
\label{figure_2}
\end{figure*}

To further explore the expansion law, we examine the geographic distribution of individual modules within the hierarchically interconnected urban system\cite{mulligan2012central,osth2021hierarchy}, which naturally results in a power-law relationship between the level of each unit and its size,  as exemplified by central place theory\cite{osth2021hierarchy}. However, due to varying standards of administrative divisions across countries, directly comparing hierarchical levels based on administrative divisions is challenging. To address this, we establish a consistent spatial hierarchy across regions by defining hierarchical levels $L$ using the H3 geospatial indexing system\cite{urberh3}. Specifically, we compute the movement matrix between H3 cells at each resolution and apply the Louvain community detection algorithm\cite{blondel2008fast} to aggregate cells, thereby defining the level units at each resolution (see Methods). Figure \ref{figure_2}a shows the resultant hierarchical partition of Boston, Massachusetts, where lower-level units cover small areas (e.g., neighborhoods), while higher-level units cover larger areas (e.g., cities and regions) and encompass lower-level units and their interconnections. 
 
As shown in Fig. \ref{figure_2}b, in the destination Boston, Massachusetts, module networks farther from home (e.g., 10km, 100km, 1000km) typically involve travel across higher spatial levels. To quantify this, each module network is aligned with spatial units across different hierarchical levels, assigning modules to a specific level, $L_c$, if their area of overlap exceeds 80\% (refer to Methods for details). Figure \ref{figure_2}c shows the distribution of $L_c$ for the corresponding module networks in Fig. \ref{figure_2}b. Analysis of Fig. \ref{figure_2}b-c indicates that module networks situated further from their home generally exhibit higher $L_c$ values on average. By analyzing all three countries’ datasets, we find that module levels and their distance from home can be well-fitted by the relationship $L_c \sim \log(d_c)$ (Fig. \ref{figure_2}d). Additionally, as shown in Fig. \ref{figure_2}e, the size $R$ of a spatial unit and its corresponding hierarchical level following $\log(R) \sim L$. Combining these findings, we establish that the module follows $\log (R) \sim \log(d_c)$. We further use the administrative hierarchy defined by the U.S. Census Bureau (county division, county, state, region) to perform a robustness check. As depicted in Fig. S12, the spatial size increases with the administrative level, adhering to the same relationship observed with the H3 delineation, and module levels increase with distance from home. In both our defined hierarchy and the administrative hierarchy, mobility modules farther from home consistently tend to align at higher levels.

In conclusion, we present a network-based segmentation of individual trajectories as a novel approach for understanding human mobility patterns across geographic extents. By examining mobility network modules in relation to distance from home, we uncover a consistent expansion phenomenon across diverse demographics and regions. This phenomenon, supported by data from the US, Senegal, and Ivory Coast, exhibits an exponent of approximately 0.6. However, unlike the US data spanning a larger geographic area, Senegal and Ivory Coast, which covering smaller regions, show reduced expansion beyond 100km, likely due to boundary effects. Additionally, the datasets from Senegal and Ivory Coast, derived from CDR, rely on cell tower distribution, resulting in sparser records in rural areas. This sparsity may introduce bias in estimating the expansion exponent. Future studies could improve accuracy by integrating additional data sources to mitigate such biases.

Furthermore, we show the connection between the expansion law and the urban hierarchical structures. Urban hierarchies facilitate mobility across different focal regions, culminating in the emergence of multiple modules in individual trajectories\cite{cabrera2023inferring}. When people are far from home, their movement tends to be at higher hierarchical levels, leading to an increased geospatial extent of the modules. Beyond hierarchy, the urban environment is also associated with the distribution of population, infrastructure, and amenities. To analyze the impact of these features, we group modules based on the number of Points of Interest (POIs) at their destinations (see Fig.S12).  Our analysis indicates that individuals visiting areas with higher POI density tend to remain within smaller module radii.  Despite this, the overall trend of expansion persists, underscoring its robustness across varying urban contexts.

This study extends prior work on the impact of home location by demonstrating that, beyond the well-established decay in travel frequency with distance\cite{fotheringham1989spatial,roy2004spatial}, individuals exhibit an expanded activity scope as distance from home increases.  This perspective and the derived expansion exponent provide valuable insights for mobility applications\cite{pappalardo2023future}. In epidemic intervention, for example, our findings can help refine epidemic models by incorporating differentiated mobility patterns based on distance from home, improving prediction accuracy and intervention effectiveness\cite{belik2011natural}. From a social equity standpoint, our results can help provide a more nuanced view of mobility segregation among racial groups, by considering neighborhood activities and movements beyond them\cite{wang2018urban}. Regarding urban resilience, this research can help cities tailor their response to emergencies and the restoration of essential services for residents and visitors at varying distances from homes\cite{haraguchi2022human}.

\clearpage
\section*{Methods}
\subsection{Data.} Our analysis of human trajectories uses three datasets (see Supplementary Text): one from the US, one from Senegal, and another from the Ivory Coast. For data preprocessing, we aggregate closely situated or overlapping visited locations (e.g., adjacent rooms within the same building) into unique hexagons using the H3 geospatial indexing system\cite{urberh3} at resolution 12, where each hexagon has an edge length of approximately 9 meters. We identify users' home locations as the hexagons they visited most frequently during nighttime hours (8 pm to 8 am).

\subsection{Extracting modules from trajectory networks.} For each user's trajectory $T=\{\theta_{1},...,\theta_{i}...\}$, where $i$ is the sequence index of the stay point $\theta_i$, we construct the trajectory network $G(T)$, with each stay point as a node and consecutive trips between two stay points as edges. To characterize $G(T)$ in geographic space, we define the edge weight $W(T)=\{w(\theta_{1},\theta_{2}),...,w(\theta_{i},\theta_{i+1}),...\}$ and $w(\theta_{i},\theta_{i+1})$ is denoted as
\begin{equation}
w(\theta_{i},\theta_{i+1})=\log(\frac{\hat{d}}{d(\theta_{i},\theta_{i+1})})
\label{weight}
\end{equation}
where $d(\theta_{i},\theta_{i+1})$ is the spatial distance between stay points $\theta_{i}$ and $\theta_{i+1}$, and $\hat{d}$ corresponds to the maximum jump distance, constrained by the geographical size of each country ($\hat{d}=4,000$km for the US, and $\hat{d}=1,000$km for Senegal and Ivory Coast). Note that $\hat{d}\geq d(\theta_{i},\theta_{i+1})$ ensures that all weights remain non-negative after the log-transformation. Equation (2) assigns greater weight to shorter distances, employing an approach known as inverse distance weighting, grounded in Tobler's first law of geography\cite{miller2004tobler}, which posits that entities in close proximity are more likely to interact than those farther apart.

Subsequently, we apply a community detection method to detect modules in the weighted directed trajectory network $G(T)$. To ensure the geographical proximity of stay points within each module, we exclude outlier staying points located farther than half the module's radius from its centroid. We also discard modules containing fewer than three locations. All characteristics of these identified modules are depicted in Fig. S2-S5.

\subsection{Delineating hierarchical levels in urban space.}  To create a consistent spatial hierarchy across countries, we construct the hierarchical levels based on the H3 geospatial indexing system\cite{urberh3}. The H3 system divides space into discrete cells, each location is assigned a cell identifier, and lower resolutions are composed of higher resolutions. Here we consider resolutions from 1 to 7, with hexagon edge lengths varying from 418 kilometers to 1 kilometer. Starting from coarsest resolution $\sigma=1$, given the spatial unit at level $L$, the hierarchy is constructed iteratively:\\
Step 1: Divide the spatial unit into cells at resolution $\sigma+1$. \\
Step 2: Aggregate cells into new spatial units at level $L-1$ using community detection on collective flow matrix between cells, defining the radius $R=\sqrt{S/\pi}$ for spatial units of size $S$.\\
The process continues through the two steps until the finest resolution, $\sigma=7$ is reached, producing a structured hierarchy where higher-level units encompass lower-level units  (Fig. \ref{figure_2}a and Fig. S13).

Given the generated spatial hierarchy, module $c$ is assigned to level $L_c=L$ if  $L$ is the smallest level where there exists a spatial unit satisfying $\frac{N_c \cap N_L}{N_c} \geq 0.8$, ensuring that at least 80\% of the module's stay points ($N_c$) are covered by the spatial unit at level $L$ ($N_L$).

\clearpage
\section*{References}

\begin{thebibliography}{10}
\expandafter\ifx\csname url\endcsname\relax
  \def\url#1{\texttt{#1}}\fi
\expandafter\ifx\csname urlprefix\endcsname\relax\def\urlprefix{URL }\fi
\providecommand{\bibinfo}[2]{#2}
\providecommand{\eprint}[2][]{\url{#2}}

\bibitem{gonzalez2008understanding}
\bibinfo{author}{Gonzalez, M.~C.}, \bibinfo{author}{Hidalgo, C.~A.} \&
  \bibinfo{author}{Barabasi, A.-L.}
\newblock \bibinfo{title}{Understanding individual human mobility patterns}.
\newblock \emph{\bibinfo{journal}{Nature}} \textbf{\bibinfo{volume}{453}},
  \bibinfo{pages}{779--782} (\bibinfo{year}{2008}).

\bibitem{belik2011natural}
\bibinfo{author}{Belik, V.}, \bibinfo{author}{Geisel, T.} \&
  \bibinfo{author}{Brockmann, D.}
\newblock \bibinfo{title}{Natural human mobility patterns and spatial spread of
  infectious diseases}.
\newblock \emph{\bibinfo{journal}{Physical Review X}}
  \textbf{\bibinfo{volume}{1}}, \bibinfo{pages}{011001} (\bibinfo{year}{2011}).

\bibitem{barbosa2018human}
\bibinfo{author}{Barbosa, H.} \emph{et~al.}
\newblock \bibinfo{title}{Human mobility: Models and applications}.
\newblock \emph{\bibinfo{journal}{Physics Reports}}
  \textbf{\bibinfo{volume}{734}}, \bibinfo{pages}{1--74}
  (\bibinfo{year}{2018}).

\bibitem{fotheringham1989spatial}
\bibinfo{author}{Fotheringham, A.~S.} \& \bibinfo{author}{O'Kelly, M.~E.}
\newblock \emph{\bibinfo{title}{Spatial Interaction Models: Formulations and
  Applications}}, vol.~\bibinfo{volume}{1} (\bibinfo{publisher}{Kluwer Academic
  Publishers}, \bibinfo{year}{1989}).

\bibitem{roy2004spatial}
\bibinfo{author}{Roy, J.~R.} \& \bibinfo{author}{Thill, J.~C.}
\newblock \bibinfo{title}{Spatial interaction modelling}.
\newblock \emph{\bibinfo{journal}{Papers in Regional Science}}
  \textbf{\bibinfo{volume}{83}}, \bibinfo{pages}{339--361}
  (\bibinfo{year}{2004}).

\bibitem{bhat1999activity}
\bibinfo{author}{Bhat, C.~R.} \& \bibinfo{author}{Koppelman, F.~S.}
\newblock \bibinfo{title}{Activity-based modeling of travel demand}.
\newblock In \emph{\bibinfo{booktitle}{Handbook of Transportation Science}},
  \bibinfo{pages}{35--61} (\bibinfo{publisher}{Springer},
  \bibinfo{year}{1999}).

\bibitem{miller2021activity}
\bibinfo{author}{Miller, H.~J.}
\newblock \bibinfo{title}{Activity-based analysis}.
\newblock \emph{\bibinfo{journal}{Handbook of Regional Science}}
  \bibinfo{pages}{187--207} (\bibinfo{year}{2021}).

\bibitem{batty2006hierarchy}
\bibinfo{author}{Batty, M.}
\newblock \bibinfo{title}{Hierarchy in cities and city systems}.
\newblock \emph{\bibinfo{journal}{Hierarchy in Natural and Social Sciences}}
  \bibinfo{pages}{143--168} (\bibinfo{year}{2006}).

\bibitem{mulligan2012central}
\bibinfo{author}{Mulligan, G.~F.}, \bibinfo{author}{Partridge, M.~D.} \&
  \bibinfo{author}{Carruthers, J.~I.}
\newblock \bibinfo{title}{Central place theory and its reemergence in regional
  science}.
\newblock \emph{\bibinfo{journal}{The Annals of Regional Science}}
  \textbf{\bibinfo{volume}{48}}, \bibinfo{pages}{405--431}
  (\bibinfo{year}{2012}).

\bibitem{osth2021hierarchy}
\bibinfo{author}{{\"O}sth, J.}, \bibinfo{author}{Reggiani, A.} \&
  \bibinfo{author}{Schintler, L.~A.}
\newblock \bibinfo{title}{Hierarchy, central place theory and computational
  modelling}.
\newblock In \emph{\bibinfo{booktitle}{Handbook on Entropy, Complexity and
  Spatial Dynamics}}, \bibinfo{pages}{454--473} (\bibinfo{publisher}{Edward
  Elgar Publishing}, \bibinfo{year}{2021}).

\bibitem{van2018christaller}
\bibinfo{author}{Van~Meeteren, M.} \& \bibinfo{author}{Poorthuis, A.}
\newblock \bibinfo{title}{Christaller and “big data”: recalibrating central
  place theory via the geoweb}.
\newblock \emph{\bibinfo{journal}{Urban Geography}}
  \textbf{\bibinfo{volume}{39}}, \bibinfo{pages}{122--148}
  (\bibinfo{year}{2018}).

\bibitem{alessandretti2020scales}
\bibinfo{author}{Alessandretti, L.}, \bibinfo{author}{Aslak, U.} \&
  \bibinfo{author}{Lehmann, S.}
\newblock \bibinfo{title}{The scales of human mobility}.
\newblock \emph{\bibinfo{journal}{Nature}} \textbf{\bibinfo{volume}{587}},
  \bibinfo{pages}{402--407} (\bibinfo{year}{2020}).

\bibitem{barthelemy2011spatial}
\bibinfo{author}{Barth{\'e}lemy, M.}
\newblock \bibinfo{title}{Spatial networks}.
\newblock \emph{\bibinfo{journal}{Physics Reports}}
  \textbf{\bibinfo{volume}{499}}, \bibinfo{pages}{1--101}
  (\bibinfo{year}{2011}).

\bibitem{miller2004tobler}
\bibinfo{author}{Miller, H.~J.}
\newblock \bibinfo{title}{Tobler's first law and spatial analysis}.
\newblock \emph{\bibinfo{journal}{Annals of the Association of American
  Geographers}} \textbf{\bibinfo{volume}{94}}, \bibinfo{pages}{284--289}
  (\bibinfo{year}{2004}).

\bibitem{blondel2008fast}
\bibinfo{author}{Blondel, V.~D.}, \bibinfo{author}{Guillaume, J.-L.},
  \bibinfo{author}{Lambiotte, R.} \& \bibinfo{author}{Lefebvre, E.}
\newblock \bibinfo{title}{Fast unfolding of communities in large networks}.
\newblock \emph{\bibinfo{journal}{Journal of Statistical Mechanics: Theory and
  Experiment}} \textbf{\bibinfo{volume}{2008}}, \bibinfo{pages}{P10008}
  (\bibinfo{year}{2008}).

\bibitem{urberh3}
\bibinfo{title}{H3: Uber’s hexagonal hierarchical spatial index}.
\newblock \bibinfo{howpublished}{\url{https://www.uber.com/blog/h3/}}.
\newblock \bibinfo{note}{Accessed: 2023-09-01}.

\bibitem{cabrera2023inferring}
\bibinfo{author}{Cabrera-Arnau, C.}, \bibinfo{author}{Zhong, C.},
  \bibinfo{author}{Batty, M.}, \bibinfo{author}{Silva, R.} \&
  \bibinfo{author}{Kang, S.~M.}
\newblock \bibinfo{title}{Inferring urban polycentricity from the variability
  in human mobility patterns}.
\newblock \emph{\bibinfo{journal}{Scientific Reports}}
  \textbf{\bibinfo{volume}{13}}, \bibinfo{pages}{5751} (\bibinfo{year}{2023}).

\bibitem{pappalardo2023future}
\bibinfo{author}{Pappalardo, L.}, \bibinfo{author}{Manley, E.},
  \bibinfo{author}{Sekara, V.} \& \bibinfo{author}{Alessandretti, L.}
\newblock \bibinfo{title}{Future directions in human mobility science}.
\newblock \emph{\bibinfo{journal}{Nature Computational Science}}
  \textbf{\bibinfo{volume}{3}}, \bibinfo{pages}{588--600}
  (\bibinfo{year}{2023}).

\bibitem{wang2018urban}
\bibinfo{author}{Wang, Q.}, \bibinfo{author}{Phillips, N.~E.},
  \bibinfo{author}{Small, M.~L.} \& \bibinfo{author}{Sampson, R.~J.}
\newblock \bibinfo{title}{Urban mobility and neighborhood isolation in
  {America’s} 50 largest cities}.
\newblock \emph{\bibinfo{journal}{Proceedings of the National Academy of
  Sciences}} \textbf{\bibinfo{volume}{115}}, \bibinfo{pages}{7735--7740}
  (\bibinfo{year}{2018}).

\bibitem{haraguchi2022human}
\bibinfo{author}{Haraguchi, M.} \emph{et~al.}
\newblock \bibinfo{title}{Human mobility data and analysis for urban
  resilience: A systematic review}.
\newblock \emph{\bibinfo{journal}{Environment and Planning B: Urban Analytics
  and City Science}} \textbf{\bibinfo{volume}{49}}, \bibinfo{pages}{1507--1535}
  (\bibinfo{year}{2022}).

\end{thebibliography}

\clearpage
\begin{addendum}

\item[Data availability] The United States dataset is anonymized location-based service records provided by
Cuebiq Inc (https://www.cuebiq.com/). The Senegal and the Ivory Coast dataset is anonymized call detail records provided by the Data for Development (D4D)
Senegal/Ivory Coast Challenge. These dataset are not publicly available due to data-sharing restrictions. To analyze the influence of demographic characteristics on mobility patterns, we additionally use publicly available data from the American Community Survey (https://
www.census.gov/programs-surveys/acs).

\item[Code availability] The codes used for data processing and analysis are available at \url{https://github.com/lucinezhong/Spatial_Expansion_Human_Mobility}.

\item[Acknowledgments] We thank Jinzhu Yu for his assistance with pre-processing the Senegal dataset and fruitful discussion. J.G. and L.Z. acknowledge the support of the US National Science Foundation under Grant No. 2047488. L.D. acknowledges the support of the National Natural Science Foundation of China (Grant No. 42422110) and the Fundamental Research Funds for the Central Universities, Peking University. Q.R.W. acknowledges the support of the US National Science Foundation under Grant No. 2125326, and 2402438.

\item[Author contributions] L.Z., L.D., Q.R.W., and J.G. conceived the project and designed the experiments;  Q.R.W. collected and analyzed the raw data; L.Z., L.D., J.G., and C.S. carried out theoretical calculations and performed the experiments; all authors wrote and edited the manuscript.

\item[Competing interests] The authors declare no competing interests.

\item[Correspondence and requests for materials] should be addressed to J.G.

\end{addendum}
\newpage
\clearpage

\clearpage
\section*{Extended Data Figures}
\renewcommand\thefigure{\arabic{figure}}
\setcounter{figure}{0}
\setcounter{table}{0}

\begin{figure*}[h!]
\renewcommand\figurename{\bf{Extended Data Fig.}}
\centering
\includegraphics[scale=1.0]{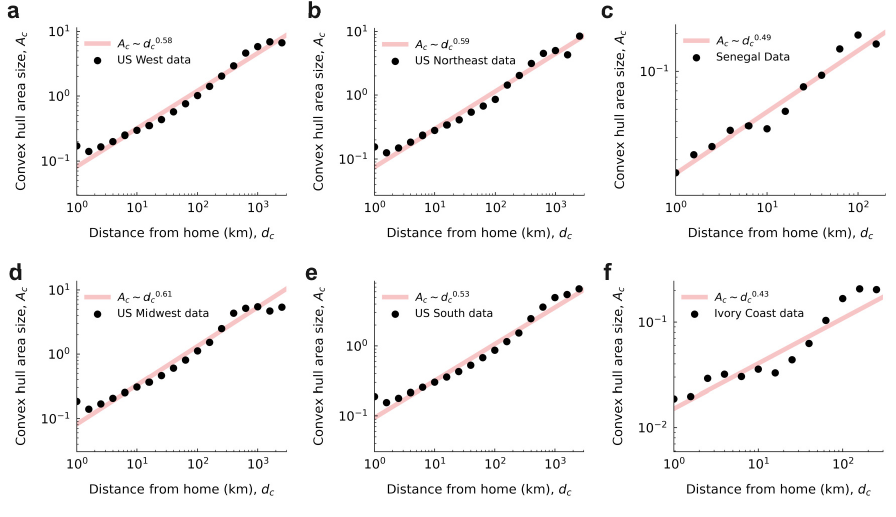}
\caption{\fontsize{10}{13.8}\linespread{1}
\selectfont{\textbf{The spatial expansion of modules regarding convex hull area size. }} Module convex hull area size, $A_c$, increases sub-linearly with its distance from home $d_c$. The exponent is around 0.55 for the U.S. data, 0.52 for the Senegal data, and 0.44 for the Ivory Coast data. }
\label{SI_dc_ac}
\end{figure*}

\clearpage
\newpage
\begin{figure*}[h!]
\renewcommand\figurename{\bf{Extended Data Fig.}}
\centering
\includegraphics[scale=1.0]{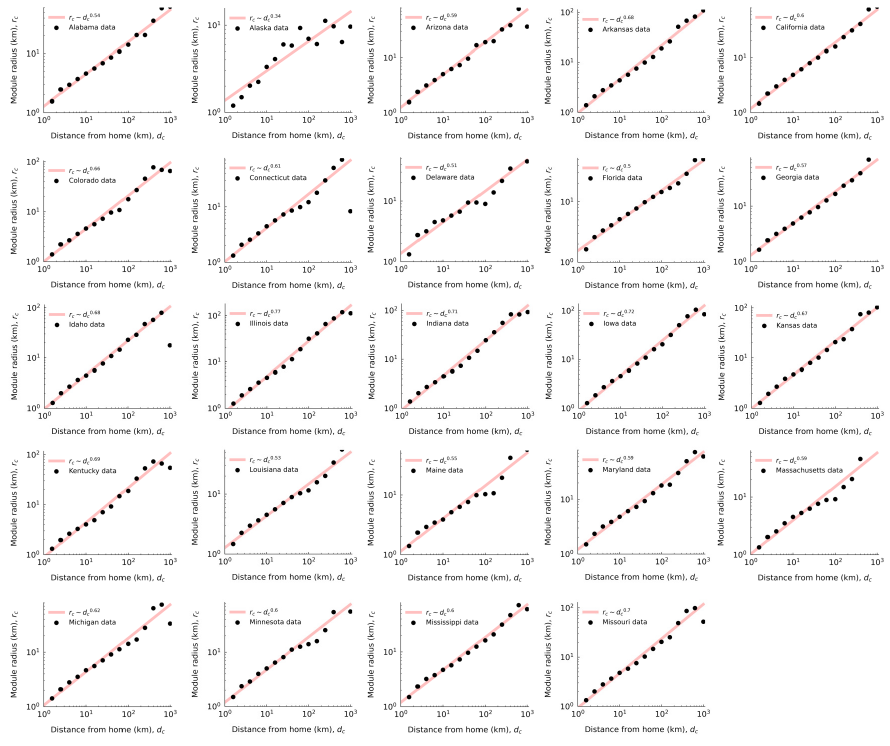}
\caption{\fontsize{10}{13.8}\linespread{1}\selectfont{\textbf{{Part-1-Module radius versus distance from home, for populations in different states.}} By categorizing users based on the states of their home locations, the spatial expansion of the module remains consistent. }} 
\label{SI_state_1}
\end{figure*}

\begin{figure*}[h!]
\renewcommand\figurename{\bf{Extended Data Fig.}}
\centering
\includegraphics[scale=1.0]{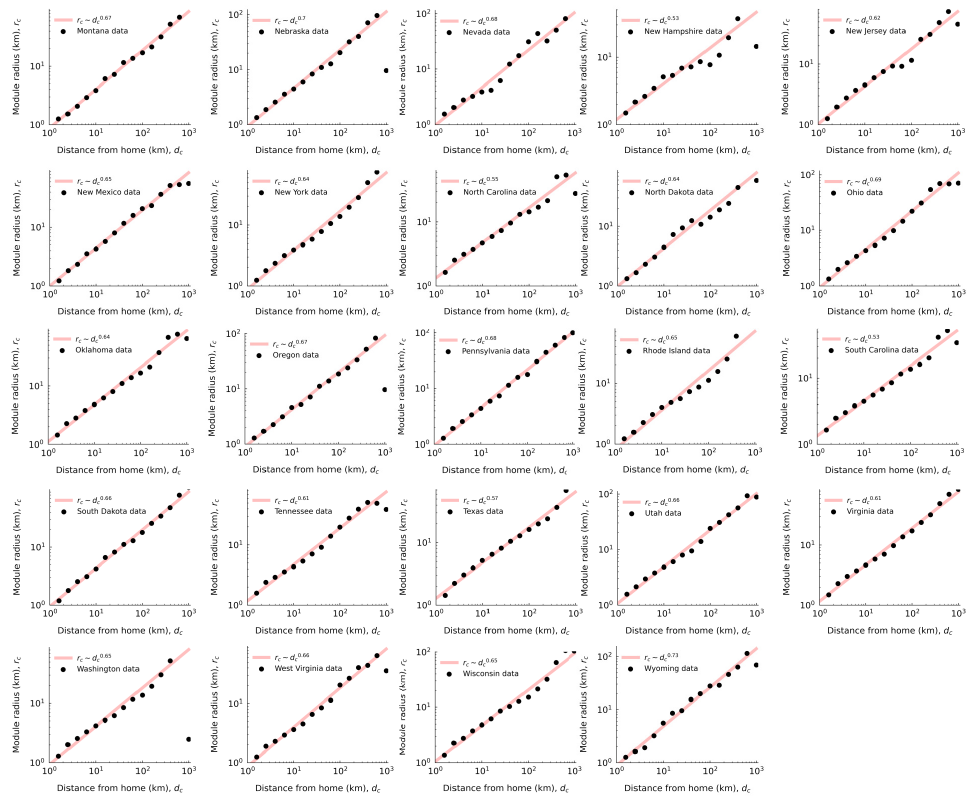}
\caption{\fontsize{10}{13.8}\linespread{1}\selectfont{\textbf{{Part-2-Module radius versus distance from home, for populations in different states.}} By categorizing users based on the states of their home locations, the spatial expansion of the module remains consistent. }} 
\label{SI_state_2}
\end{figure*}

\newpage
\begin{figure*}[h!]
\renewcommand\figurename{\bf{Extended Data Fig.}}
\centering
\includegraphics[scale=1]{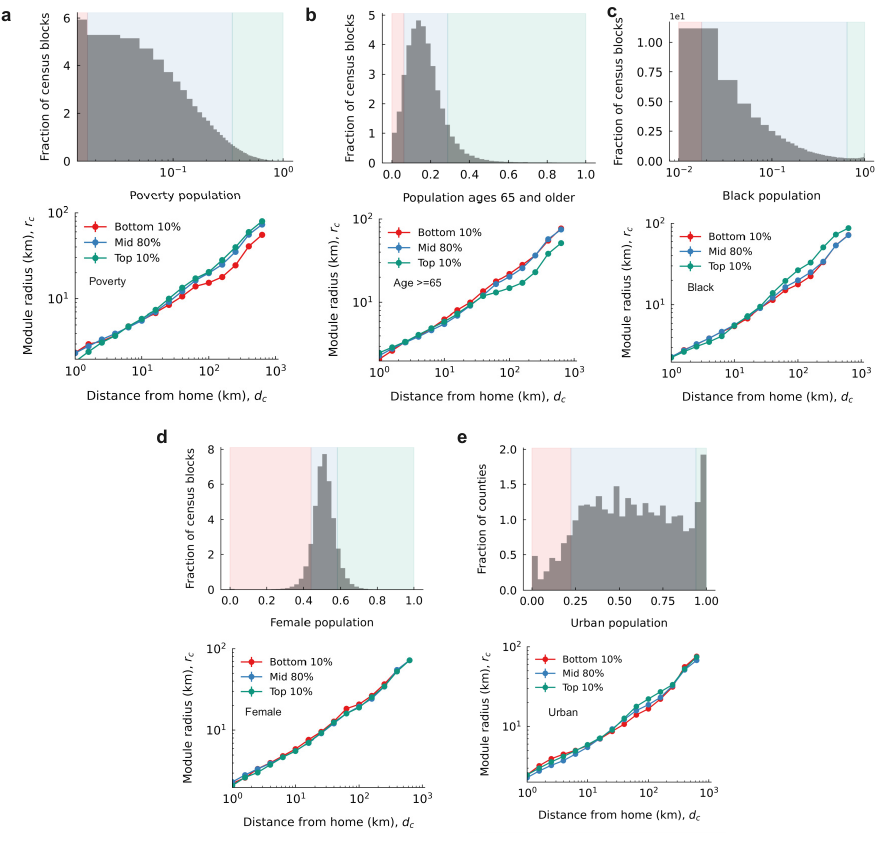}
\caption{\fontsize{10}{13.8}\linespread{1}\selectfont{\textbf{Module radius versus distance from home, for populations in different demographic attributes.}}  By categorizing users based on the proportions of the poverty population in their home locations \textbf{(a)}, the elderly population (age 65 and older) \textbf{(b)}, the female population \textbf{(c)},  the black population \textbf{(d)}, the urban  population  \textbf{(e)}, the spatial expansion of module remains consistent across various user groups.}
\label{SI_demographic}
\end{figure*}

\end{document}